\documentclass[]{bmcart}

\usepackage{amsthm,amsmath}
\RequirePackage{natbib}
\usepackage{bm}
\usepackage[utf8]{inputenc} 
\usepackage{graphicx}
\graphicspath{{./Figs/}}
\usepackage{cancel}
\usepackage{booktabs}

\startlocaldefs

\newcommand{\Bb}{{\boldsymbol{\mathnormal b}}}

\newcommand{\Be}{{\boldsymbol{\mathnormal e}}}

\newcommand{\Bk}{{\boldsymbol{\mathnormal k}}}

\newcommand{\Bn}{{\boldsymbol{\mathnormal n}}}

\newcommand{\Br}{{\boldsymbol{\mathnormal r}}}
\newcommand{\Bs}{{\pmb{\mathnormal s}}}
\newcommand{\Bt}{{\boldsymbol{\mathnormal t}}}

\newcommand{\Bv}{{\boldsymbol{\mathnormal v}}}

\newcommand{\CC}{{\boldsymbol{\mathnormal C}}}

\newcommand{\BE}{{\boldsymbol{\mathnormal E}}}
\newcommand{\BF}{{\boldsymbol{\mathnormal F}}}

\newcommand{\BI}{{\boldsymbol{\mathnormal I}}}

\newcommand{\BL}{{\boldsymbol{\mathnormal L}}}

\newcommand{\BS}{{\boldsymbol{\mathnormal S}}}

\newcommand{\superscr}[1]{\ensuremath{{}^{\rm #1}}}
\newcommand{\subscr  }[1]{\ensuremath{{}_{\rm #1}}}

\newcommand{\Balpha }{\ensuremath{\boldsymbol\alpha}}
\newcommand{\Bbeta  }{\ensuremath{\boldsymbol\beta}}

\newcommand{\Brho}{{\boldsymbol{\rho}}}

\newcommand{\Bve    }{\ensuremath{\boldsymbol\varepsilon}}

\newcommand{\vp     }{\ensuremath{\varphi}}
\newcommand{\Bbetapl}{\ensuremath{\boldsymbol\beta\superscr{pl}}}

\newcommand{\taub   }{\ensuremath{           \tau \subscr{b}}}

\newcommand{\dy} {{\rm d}y}

\newcommand{\figref}[1]{Fig.~\ref{#1}}

\newcommand \MZ [1] {\bgroup\noindent[\textcolor{blue}{\textbf{MZ}: #1}]\egroup\ignorespacesafterend}

\newcommand \Mdel [1] {\bgroup\noindent[\textcolor{red}{\textbf{Mdel}: #1}]\egroup\ignorespacesafterend}
\endlocaldefs

\newcommand \Madd [1] {\bgroup\noindent[\textcolor{blue}{\textbf{Madd}: #1}]\egroup\ignorespacesafterend}
\endlocaldefs

\begin{document}

\begin{frontmatter}

\begin{fmbox}
\dochead{Research}


\title{A computationally efficient implementation of continuum dislocation dynamics: Formulation and application to ultrafine-grained Mg polycrystals}


\author[
   addressref={aff2},                   
   corref={aff1},                       
   email={luoxilucy@126.com}   
]{\inits{XL}\fnm{Xi} \snm{Luo}}
\author[
   addressref={aff1,aff2},
   email={michael.zaiser@fau.de}
]{\inits{MZ}\fnm{Michael} \snm{Zaiser}}


\address[id=aff1]{
  \orgname{Institute of Materials Simulation (WW8), Friedrich-Alexander University Erlangen-N\"urnberg (FAU)}, 
  \street{Dr.-Mack-Strasse 77},                     %
  \postcode{90762}                                
  \city{F\"urth},                              
  \cny{Germany}                                    
}
\address[id=aff2]{
  \orgname{Applied Mechanics and Structure Safety Key Laboratory of Sichuan Province, School of Mechanics and Aerospace Engineering}, 
  \city{Chengdu}
  \cny{P.R. China}                                    
}


\begin{artnotes}
\end{artnotes}

\end{fmbox}


\begin{abstractbox}

\begin{abstract} 
Continuum dislocation dynamics (CDD) represents the evolution of systems of curved and connected dislocation lines in terms of density-like field variables which include the volume density of loops (or 'curvature density') as an additional field. Since dislocation curvature represents a spatial derivative of the underlying discrete dislocation density tensor, the curvature field evolution equation of necessity contains numerically inconvenient higher-order derivatives of the density fields. We propose a simple approximation to express curvature in terms of density fields, and demonstrate its application to a benchmark problem in deformation of Mg polycrystals. 
\end{abstract}




\end{abstractbox}
%

\end{frontmatter}



\section[Introduction]{Introduction}
All continuum theories of dislocation motion in one way or another relate to the classical work of Kr\"oner \cite{Kroener58_Book} and Nye \cite{Nye53_AM} who describe the dislocation system in terms of a rank-2 tensor field $\Balpha$ defined as the curl of the plastic distortion, $\Balpha = - \nabla \times \Bbetapl$. The rate of the plastic distortion due to the evolution of the dislocation density tensor reads $\partial_t \Bbetapl=-\Bv \times \Balpha$ where the dislocation velocity vector $\Bv$ is defined on the dislocation lines, hence \cite{Mura63_PM}
\begin{align}
\partial_t \Balpha = - \nabla \times [\Bv \times \Balpha].
\end{align}
The fundamental setting provided by the classical continuum theory of dislocation systems has inspired many models (e.g. \cite{Sedlacek03_PM, Acharya06_JMPS,Xiang09_JMPS, Zhu15_JMPS}). Irrespective of the specific formulation, a main characteristic of the CCT is that, in each elementary volume, the dislocation tensor can measure only the minimum amount of dislocations which are necessary for geometrical compatibility of plastic distortion (`geometrically necessary' dislocations). The information contained in the classical theory is thus only complete if a very high spatial resolution is employed such that no 'redundant' dislocations of zero net Burgers vector are contained in the elementary volumes. 

A further observation is that a thermodynamically consistent formulation of dislocation motion should account for the thermodynamic driving forces which, for dislocation glide motion, are provided by the resolved shear stresses on the respective slip systems. Because of this, it is important to resolve the dislocation density tensor into slip system specific contributions whose velocities can be related to the slip system specific driving forces -- in other words, it is desirable to use a crystal plasticity formulation. This was done in the works of El-Azab and co-workers \citep{Xia15_MSMSE, Xia16_MSMSE,Lin2020implementation} and has shown promising results in modelling dislocation pattern formation. The tensor $\Balpha$ is decomposed into contributions of dislocations from the different slip systems $\varsigma$ in the form $\Balpha = \sum_{\varsigma}\Brho^{\varsigma}\otimes\Bb^{\varsigma}$ where $\Bb^{\varsigma}$ is the Burgers vector of dislocations on slip system $\varsigma$ and the dislocation density vector $\Brho^{\varsigma}$ of these dislocations points in their local line direction. Accordingly, the evolution of the dislocation density tensor is written as $\partial_t \Balpha = \sum_{\varsigma}\partial_t \Brho^{\varsigma}\otimes\Bb^{\varsigma}$ with $\partial_t \Brho^{\varsigma} = \nabla \times [\Bv^{\varsigma} \times \Brho^{\varsigma}]$ where the dislocation velocities $\Bv^{\varsigma}$ are again slip system specific.  We use a description of the dislocation system by slip system specific dislocation density vectors as the starting point of our subsequent discussion. This description is complete as long as each elementary volume contains, for a given slip system, only dislocations of the same orientation ('single-valued dislocation density fields'), in which case it provides a complete and kinematically exact plasticity theory.

Moving from the micro- to the macroscale requires the use of elementary volumes that contain dislocations of multiple orientations. Averaging operations are then needed which lead to the presence of 'redundant' dislocations that do not contribute to the average dislocation density vectors, which only describe the so-called 'geometrically necessary' dislocations that remain, after averaging, on the given scale of resolution. From a point of view of dislocation kinematics the problem of these 'redundant' dislocations is that they have no unique orientation, and hence no unique direction of motion. In simple words, we do not know where they are going.

Some continuum theories try to resolve this averaging problem by describing the microstructure by multiple dislocation density fields which each represent a specific dislocation orientation $\vp$ on a slip system $\varsigma$. Accordingly, all dislocations of such a partial population move in the same direction in such a way that $\langle\partial_t \Brho_{\vp}^{\varsigma}\rangle \approx \nabla \times [\langle\Bv_{\vp}^{\varsigma}\rangle \times \langle\rho_{\vp}^{\varsigma}\rangle]$. Along this line Groma, Zaiser and co-workers \citep{Groma03_AM,groma2016dislocation,wu2018instability,wu2021cell} developed  statistical approaches for evolution of 2D systems of straight, positive and negative edge dislocations and analyzed the associated patterning phenomena. Other authors  \cite{Arsenlis04_JMPS,Reuber14_AM,Leung15_MSMSE} developed 3D models by considering additional orientations, e.g. edge vs screw dislocations. However, extending the approach to 3D systems where connected and curved dislocation lines can move perpendicular to their line direction while remaining topologically connected is not straightforward, and most models use simplified kinematic rules for coupling the motion of dislocations of different orientations that cannot in general guarantee  dislocation connectivity (see Ref. \cite{Monavari16_JMPS} for discussion).  

A third line of reasoning starts from the idea of evolving a continuous orientation distribution of dislocation lines in each point of space, leading to a higher dimensional phase space where densities carry additional information about their line orientation and curvature in terms of continuous orientation variables $\vp$ \citep{Hochrainer07_PM,Zaiser07_PM}. In this phase space, the microstructure is described by dislocation orientation distribution functions (DODF). Since this approach is numerically challenging \citep{sandfeld2010numerical}, it has been proposed to approximate the evolution of the DODF in terms of its alignment tensor expansion  \cite{Hochrainer15_PM,Monavari16_JMPS}. The components of the dislocation density alignment tensors can be envisaged as density-like fields which contain more and more detailed information about the dislocation orientation distribution. This approach has been used to simulate various phenomena including dislocation patterning \cite{Sandfeld15_MSMSEa} and co-evolution of phase and dislocation microstructure \citep{Wu17_IJP}. The formulation in terms of alignment tensors has proven versatile since one can expand the elastic energy functional of the dislocation system in terms of dislocation density alignment tensors \cite{Zaiser15_PRB} and then use this functional to derive the dislocation velocity in a thermodynamically consistent manner \cite{Hochrainer16_JMPS}. While originally formulated as a continuum theory of dislocation transport, the processes of dislocation annihilation and multiplication were introduced into the approach  \cite{monavari2018annihilation,sudmanns2019dislocation}. Numerical implementations were provided in the DAMASK crystal plasticity framework \cite{roters2019damask}.

Despite its attractive features, the practical implementation of alignment tensor based continuum dislocation dynamics has suffered from an important drawback: Besides the dislocation density alignment tensors, the theory necessarily requires for kinematic consistency an additional field variable which can be understood as a dislocation curvature density, or equivalently a density of (partial) dislocation loops. The evolution equation for this variable contains second-order derivatives of the dislocation fluxes, which in turn 
may depend on derivatives of the dislocation alignment tensors \citep{Hochrainer16_JMPS}, leading to high-order spatial derivatives which are numerically awkward to handle. 

Here we propose a simple remedy that allows, by incurring a mild approximation in the averaging procedure, to express the curvature density in terms of the dislocation density fields, yielding a closed description of dislocation transport. We illustrate the performance of the method for a few limiting cases as well as for a full-field simulation of deformation of Mg polycrystals. 

\section[CDD]{Theoretical background}

We describe the evolution of the dislocation microstructure in a material reference system. 
Dislocations of Burgers vectors $\Bb^{\varsigma}$ are assumed to move only by glide (unless stated otherwise) and are therefore confined to their slip planes with slip plane normal vectors $\Bn^{\varsigma}$. This motion generates a plastic shear $\gamma^{\varsigma}$ in the direction of the unit slip vector $\Bs^{\varsigma}=\Bb^{\varsigma}/b^{\varsigma}$ where $b^{\varsigma}$ is the modulus of $\Bb^{\varsigma}$. We use the following sign convention: A dislocation loop which expands under positive resolved shear stress  is called a positive loop, the corresponding dislocation density vector $\Brho^{\varsigma}$ points in counter-clockwise direction with respect to the slip plane normal $\Bn^{\varsigma}$. Summing the plastic shear tensors of all slip systems gives the plastic distortion in the material frame: $\Bbetapl=\sum_{\varsigma}\gamma^{\varsigma} \Bn^{\varsigma} \otimes \Bb^{\varsigma}/b^{\varsigma}$. 
A slip system specific Levi-Civita tensor $\Bve^{\varsigma}$ with coordinates $\varepsilon_{ij}^{\varsigma}$ is constructed by contracting the fully antisymmetric Levi-Civita operator with the slip plane normal, $\varepsilon_{ij}^{\varsigma}=\varepsilon_{ijk}n_k^{\varsigma}$. The operation $\Bt.\Bve^{\varsigma} =: \Bt_{\perp}$ then rotates a vector $\Bt$ on the slip plane counter-clockwise by $90^\circ$ around $\Bn^{\varsigma}$. 

The quantity which is fundamental to density based crystal plasticity models is the slip system specific dislocation density vector $\Brho^{\varsigma}$. The modulus of this vector defines a scalar density $\rho^{\varsigma} = |\Brho^{\varsigma}|$ and the unit tangent vector $\Bt^{\varsigma} = \Brho^{\varsigma}/\rho^{\varsigma}$ gives the local dislocation direction. 

To derive evolution equations, we start from the slip system specific Mura equation in the form 
\begin{equation}
\partial_t \Brho^{\varsigma} = \nabla \times [\Bv^{\varsigma} \times \Brho^{\varsigma}].
\label{eq:Mura1}
\end{equation}
where we assume that the spatial resolution is sufficiently high such that the dislocation line orientation $\Bt$ is uniquely defined in each spatial point. If deformation occurs by crystallographic slip, then the dislocation velocity vector must in this case have the local direction $\Be_v = \Bt \times \Bn^{\varsigma} = \Brho^{\varsigma}\times \Bn^{\varsigma}/\rho^{\varsigma}$. We obtain
\begin{equation}
\partial_t \Brho^{\varsigma} = \nabla \times [\Bt \times \Bn^{\varsigma} \times \Brho^{\varsigma} v^{\varsigma}] = \nabla \times [\frac{\Brho^{\varsigma} \times \Bn^{\varsigma} \times \Brho^{\varsigma}}{\rho^{\varsigma}} v^{\varsigma}].
\label{eq:Mura2}
\end{equation}
where the velocity magnitude $v^{\varsigma}$ depends on the local resolved shear stress in the respective slip system. This equation is {\em kinematically non-linear} which makes it equation difficult to average. To obtain an equation which is linear in a dislocation density variable (and therefore can be averaged straightforwardly) we use $\Brho^{\varsigma} = \Bt \rho^{\varsigma}$ and $\nabla \times \Bt \times \Bn^{\varsigma} \times \Bt = \nabla .\Bve^{\varsigma}$. This gives the simple result
\begin{eqnarray}
\partial_t \Brho^{\varsigma} = \Bve^{\varsigma}.\nabla(\rho^{\varsigma} v).
\label{eq:Murarewrite}
\end{eqnarray}
Note that, as a corollary, we can recover the Orowan eqation for the considered slip system by observing that $\partial_t \Balpha^{\varsigma} = \partial_t \Brho^{\varsigma} \otimes \Bb^{\varsigma} = - \partial_t \Bbeta^{{\rm pl},\varsigma}$ that the plastic strain rate and the shear strain rate on the considered slip system fulfil the Orowan equation 
\begin{equation}
\partial_t \Bbeta^{{\rm pl},\varsigma} = [\Bn^{\varsigma} \otimes \Bb^{\varsigma}] \rho^{\varsigma} v= [\Bn^{\varsigma} \otimes \Bs^{\varsigma}] \partial_t \gamma^{\varsigma} \quad,\quad
\partial_t \gamma^{\varsigma} = \rho^{\varsigma} b^{\varsigma} v.
\label{eq:Strainrate}
\end{equation}
We now need to derive an equation for the scalar density $\rho^{\varsigma}$. This is straightforward: we use that $(\rho^{\varsigma})^2 = \Brho^{\varsigma}.\Brho^{\varsigma}$, hence 
\begin{equation}
\partial_t \rho^{\varsigma} = \frac{\Brho^{\varsigma}}{\rho^{\varsigma}} \cdot \partial_t \Brho^{\varsigma} = \frac{\Brho^{\varsigma}}{\rho^{\varsigma}} \cdot  \Bve^{\varsigma}.\nabla(\rho^{\varsigma} v).
\end{equation}
This equation is still exact in case of single-valued dislocation density fields, for which it is simply equivalent to Eq. (2). To proceed further, \cite{monavari2018annihilation} re-wrote this equation as  
\begin{equation}
\partial_t \rho^{\varsigma} = \nabla \cdot (\Bve^{\varsigma} \cdot \Brho^{\varsigma} v^{\varsigma}) + q^{\varsigma} v^{\varsigma}
\label{eq:rho}
\end{equation}
where the quantity 
\begin{equation}
q^{\varsigma} := \rho^{\varsigma} (\Bve^{\varsigma}.\nabla\cdot \Bt) .
\label{eq:qmicro}
\end{equation}
is the product of the local dislocation density and the curvature of the unit vector field $\Bt = \Brho^{\varsigma}/\rho^{\varsigma}$. In the classical treatment of CDD \cite{Hochrainer15_PM,monavari2018annihilation}, the $q^{\varsigma}$ ('curvature density fields') are  treated as an independent field variables whose evolution is governed by transport equations that contain higher-order alignment tensors, ultimately leading to an infinite hierarchy of equations. The advantage of this hierarchy is that the equations are linear in the field variables (alignment tensors) and thus amenable to averaging. The disadvantage is that the equations still are in need of closure (see \cite{Monavari16_JMPS} for an extensive discussion of this point). Besides, they are numerically unpleasant. 

To avoid this problem, we depart from the treatment of Monavari and Zaiser \cite{monavari2018annihilation} and take Eq. (\ref{eq:qmicro}) as our starting point. Upon averaging over a volume containing dislocations of different orientation and denoting the averaging operation by $\langle \dots \rangle$, we can formally write
\begin{equation}
	\langle q \rangle = \langle \rho^{\varsigma} \Bve^{\varsigma}\cdot\nabla\cdot \Bt\rangle
	=  \langle \rho^{\varsigma} \rangle \Bve^{\varsigma} \cdot\nabla\cdot \left(\frac{\langle\Brho^{\varsigma}\rangle}{\langle\rho^{\varsigma}\rangle} \right) + q^{\varsigma}_{\rm H} .
	\label{eq:qav}
\end{equation}
The first term on the right-hand side can be understood as the product of the average density and the derivative of an average unit tangent vector, multiplied with the fraction of geometrically necessary dislocations: 
\begin{equation}
	\langle \rho^{\varsigma} \rangle \Bve^{\varsigma} \cdot\nabla\cdot \left(\frac{\langle\Brho^{\varsigma}\rangle}{\langle\rho^{\varsigma}\rangle} \right)
	= \langle \rho^{\varsigma} \rangle \Bve^{\varsigma} \cdot\nabla\cdot f_{\rm GND}^{\varsigma} \langle \Bt \rangle =: q_{\rm GND}^{\varsigma} ,
	\label{eq:qgnd}
\end{equation}
where the GND fraction of dislocations is defined as $f_{\rm GND}^{\varsigma} = 
|\langle\Brho^{\varsigma}\rangle|/\langle\rho^{\varsigma}\rangle$. Thus, the physical interpretation of the first term of the right-hand side is that of a {\em mean GND curvature}. For a statistically homogeneous dislocation arrangement, this term vanishes. Since the dislocations may form loops even in a statistically homogeneous dislocation arrangement, their curvature is in general not zero. Therefore, the residual $q^{\varsigma}_{\rm H}$, which can be envisaged as a homogeneous loop density, emerges in Eq. (\ref{eq:qav}) upon averaging. On the other hand, for a single-valued dislocation density field, $\rho^{\varsigma} = \langle\rho^{\varsigma}\rangle, \Bt = \langle \Bt \rangle, f_{\rm GND}^{\varsigma} =1$ and therefore the residual term $q^{\varsigma}_{\rm H}$ vanishes. We may thus interpret $q^{\varsigma}_{\rm H}$ as the loop density of the 'redundant' or 'statistically stored' dislocations. 

The evolution of the averaged dislocation density is then given by
\begin{equation}
	\partial_t \langle \rho^{\varsigma} \rangle = \frac{\langle \Brho^{\varsigma}\rangle}{\langle\rho^{\varsigma}\rangle} \cdot  \Bve^{\varsigma}\cdot\nabla(\langle \rho^{\varsigma} \rangle v^{\varsigma}) + q^{\varsigma}_{\rm H} v^{\varsigma} =  \frac{\langle \Brho^{\varsigma}\rangle}{\langle\rho^{\varsigma}\rangle} \cdot  \Bve^{\varsigma}.\nabla(\langle \rho^{\varsigma} \rangle v) + q^{\varsigma}_{\rm H} v^{\varsigma} 
	\label{eq:transport}
\end{equation}
We take this equation as the starting point of our considerations. In the following, all densities are understood as averages over volumes containing dislocations of multiple orientations, and accordingly we simplify notations by omitting the averaging brackets. 
To arrive at a closed set of equations, the loop densities $q^{\varsigma}_{\rm H}$ need to be related to the dislocation densities $\rho^{\varsigma}$ and dislocation density vectors $\Brho^{\varsigma}$. While our considerations until now have been exclusively related to dislocation kinematics, this requires us to make constitutive assumptions, i.e., to formulate a model.

\section{Model formulation}

On each slip system $\varsigma$, we consider the scalar total dislocation density $\rho^{\varsigma}$ and dislocation density vector $\Brho^{\varsigma}$. The magnitude of which may be interpreted as a scalar GND density and is denoted as $\rho^{\varsigma}_{\rm GND} = |\Brho^{\varsigma}|$ . 

\subsection{Dislocation transport and dislocation generation}

As discussed in detail in the previous section, we describe dislocation transport for a given slip system, by two coupled equations for the total dislocation density and for the dislocation density vector. Since transport of curved dislocations is necessarily accompanied by changes in line length, these equations contain curvature or loop densities:
\begin{eqnarray}
	\partial_t \Brho^{\varsigma}_{\rm TR} &=& \Bve^{\varsigma}\cdot \nabla (\rho^{\varsigma}v^{\varsigma}),\\
	\partial_t \rho^{\varsigma}_{\rm TR} &=& \frac{\Brho^{\varsigma}}{\rho^{\varsigma}}\cdot \Bve^{\varsigma}\cdot \nabla (\rho^{\varsigma}v^{\varsigma}) + q^{\varsigma}_{\rm H} v^{\varsigma} 
\end{eqnarray}
An equivalent form of these equations is 
\begin{eqnarray}
	\partial_t \Brho^{\varsigma}_{\rm TR} &=& \Bve^{\varsigma}\cdot \nabla (\rho^{\varsigma}v^{\varsigma}),\\
	\partial_t \rho^{\varsigma}_{\rm TR} &=&  \Bve^{\varsigma}\cdot \nabla (\Brho^{\varsigma}v^{\varsigma}) + (q^{\varsigma}_{\rm GND} + q^{\varsigma}_{\rm H}) v^{\varsigma}.
	\label{eq:density}
\end{eqnarray}
where the GND curvature $q^{\varsigma}_{\rm GND}$ is given by Eq. (\ref{eq:qgnd}). This second version of the transport equations demonstrates that the effect of GND curvature is akin to additional source terms in the equations for the total dislocation densities $\rho^{\varsigma}$. The second curvature term $q^{\varsigma}_{\rm H}$ which describes the dislocation loop density in a homogeneous dislocation arrangement and must be specified constitutively. In absence of dislocation transport ('zeroth-order continuum dislocation dynamics', \cite{monavari2018annihilation}), only this term is present. 

Monavari and Zaiser \cite{monavari2018annihilation} formulate evolution equations for the dislocation loop densities that account for the nucleation and recombination (i.e., partial annihilation) of loops. For an evolving homogeneous dislocation system, these authors showed that, after a short initial transient, the average loop radius is proportional to the dislocation spacing, in line with generic scaling properties of dislocation systems \cite{zaiser2014scaling}. Also, only loops that expand under the applied stress survive. This implies that the homogeneous loop density scales like
\begin{equation}
	q^{\varsigma}_{\rm H} = \eta_{\rm L} (\rho^{\varsigma})^{3/2} {\rm sign}(v^{\varsigma}). 
	\label{eq:qhom}
\end{equation}
Neglecting transients (which may however be important e.g. upon changes in loading path), we use this equation to describe dislocation density increase due to loop expansion. 

\subsection{Dislocation annihilation}

Dislocation annihilation was analyzed in Ref. \cite{monavari2018annihilation} in terms of the partial recombination of dislocation loops expanding on closely adjacent slip planes. The recombination rate in general depends on the orientation distribution of dislocations, and on the orientation dependent annihilation distance. The resulting annihilation rates for the dislocation density can be written in the form (note that the dislocation density vectors are not changed by annihilation processes)
\begin{equation}
	\partial_t \rho^{\varsigma}_{\rm A} = - 2 y_{\rm A}\phi(f_{\rm GND}^{\varsigma}) (\rho^{\varsigma})^2 v^{\varsigma}
\end{equation}
where $y_{\rm A}$ is an angle-averaged annihilation distance. The function $\phi(f)$ describes the dependency of the annihilation rate on the GND fraction of the dislocation density, it has the general properties $\phi(0) = 1, \phi'(0) = 0$ and $\phi(1) = 0$. The specific functional form of $\eta$ depends on the angular distribution of dislocations and on the angle dependence of the annihilation distance. Several cases have been discussed by \citet{monavari2018annihilation}, here we consider the simplest functional form which is given by 
\begin{equation}
	\phi(f) = 1 - f^2.
\end{equation}
With these expressions, the evolution of the dislocation arrangement is given by
\begin{eqnarray}
	\partial_t \Brho^{\varsigma} &=& \Bve^{\varsigma}\cdot \nabla (\rho^{\varsigma}v^{\varsigma}),
	\label{eq:Brho}\\
	\partial_t \rho^{\varsigma} &=& \frac{\Brho^{\varsigma}}{\rho^{\varsigma}}\cdot \Bve^{\varsigma}\cdot \nabla (\rho^{\varsigma}v^{\varsigma}) + \eta_{\rm L} (\rho^{\varsigma})^{3/2} |v|^{\varsigma} - 2[(\rho^{\varsigma})^2-|\Brho^{\varsigma}|^2] y_{\rm A} v^{\varsigma}.
	\label{eq:rho}	
\end{eqnarray}
To close the model equations, we still need to specify the dislocation velocity. This is discussed in the following section.

\subsection{Dislocation velocities} 

We assume the scalar dislocation glide velocities $v^{\varsigma}$ to be governed by thermally activated processes, which leads to a dependency of the glide velocities on the resolved shear stresses of the form
\begin{equation}
	v^{\varsigma} = v_0 \exp\left(-\frac{Q}{kT}\right) \sinh \left(\frac{V_{\rm a}}{kT} \tau_{\rm eff}^{\varsigma}\right)
	\label{eq:v}
\end{equation}
Here, $v_0$ is a pre-factor with the dimension of a velocity, $Q$ is an activation enthalpy, $T$ temperature, and $V_{\rm a}$ an activation volume. The effective shear stress $\tau_{\rm eff}^{\varsigma}$ provides the effective driving force for dislocation motion. As discussed 
in Ref. \cite{wu2022thermodynamic}, thermodynamic considerations impose some constraints on the definition of this driving force which is not independent of the considered kinematics: the generation of dislocation line length during dislocation motion increases the defect energy of the crystal, and the corresponding driving force must be defined in such a manner as to ensure that the expended mechanical power exceeds the increase in defect energy everywhere in the system. 

To discuss the implications of this idea, we look at Eq. (\ref{eq:density}) and observe that the first term on the right-hand side is conservative while the second term provides a net line length change of $(q^{\varsigma}_{\rm GND} + q^{\varsigma}_{\rm H}) v^{\varsigma}$. With a dislocation line energy which we take for simplicity as $e_{\rm D} = \eta_{\rm D}\mu b^2$, this corresponds to a defect energy power $P_{\rm D}^{\varsigma} = (q^{\varsigma}_{\rm GND} + q^{\varsigma}_{\rm H}) \eta_{\rm D}\mu b^2 v^{\varsigma}$ which must not exceed the mechanical power expended in the same slip system, $P_{\rm D}^{\varsigma} \le P_{\rm M}^{\varsigma} = \tau^{\varsigma} \dot{\gamma}^{\varsigma}$. With Eqs. (\ref{eq:qgnd}) and (\ref{eq:qhom}), and using $\dot{\gamma}^{\varsigma} = \rho^{\varsigma}b^{\varsigma}v^{\varsigma}$, this leads to the inequality
\begin{equation}
	\tau^{\varsigma}v^{\varsigma} \ge \eta_{\rm D}\mu b^{\varsigma}v^{\varsigma}  \left[  \Bve^{\varsigma} \cdot\nabla\cdot \left(\frac{\Brho^{\varsigma}}{\rho^{\varsigma}} \right) + {\rm sign}(v^{\varsigma}) \eta_{\rm L} \mu b^{\varsigma} \sqrt{\rho^{\varsigma}} \right]. 
	\label{eq:ineq}
\end{equation}
We satisfy this inequality by defining, in Eq. (\ref{eq:v}), the effective resolved shear stress which provides the thermodynamic driving force for dislocation motion as
\begin{equation}
	\tau_{\rm eff}^{\varsigma} = \left\{ \begin{array}{l}
	\tau^{\varsigma} + \taub^{\varsigma} - (\tau^{\varsigma}_{\rm f}+\tau^{\varsigma}_{0}){\rm sign}(\tau^{\varsigma}) \quad,\quad 
	|\tau^{\varsigma} + \taub^{\varsigma}| > \tau^{\varsigma}_{\rm f} + \tau^{\varsigma}_{0} \\
	0 \quad,\quad 
	|\tau^{\varsigma} + \taub^{\varsigma}| \le \tau^{\varsigma}_{\rm f} + \tau^{\varsigma}_{0} .
	\end{array}\right.
\end{equation}
Here, the 'back stress' related to GND storage is given by 
\begin{equation}
	\taub^{\varsigma} = \eta_{\rm D}  \mu b^{\varsigma}  \Bve^{\varsigma} \cdot\nabla\cdot \left(\frac{\Brho^{\varsigma}}{\rho^{\varsigma}} \right).
	\label{eq:taub}
\end{equation}
The critical resolved shear stress required to overcome the dislocation-lattice interaction is denoted by $\tau^{\varsigma}_{0}$, and the friction-like stress $\tau^{\varsigma}_{\rm f}$ is assumed in the Taylor-like form
\begin{equation}
\tau^{\varsigma}_{\rm f} = \mu b^{\varsigma} \sqrt{\sum_{\varsigma'} h_{\varsigma\varsigma'} \rho^{\varsigma'}} 
\end{equation}
where the positively definite coefficients $h_{\varsigma\varsigma'}$ define a latent hardening matrix and the inequality costraint Eq. (\ref{eq:ineq}) imposes the condition $h_{\varsigma\varsigma'} > \eta_{\rm D}^2 \eta_{\rm L}^2$. 

\section{Numerical implementation}

The constitutive model is implemented within the Dusseldorf Advanced Material Simulation Kit, DAMASK \cite{roters2012damask}. The governing boundary value problem for mechanical equilibrium is solved by a Fast Fourier Transform (FFT) based spectral method, a detailed description of the spectral solver has been given by Roters et al. \cite{roters2019damask}. Here, we only give a brief outline and then focus on asptcts of the time integration procedure that concern the spatio-temporal dynamics of dislocation densities.

\subsection{Elasticity and plasticity}

Our problem is composed of two complementary parts, namely a kinematic evolution of the dislocation system coupled with a solution of the mechanical boundary value problem. We first describe the elastic-plastic boundary value problem which is formulate in terms of the second Piola–Kirchhoff stress tensor $\BS$ that relates to its work conjugated Green-Lagrange strain tensor $\BE$ via
\begin{equation}
	\BS = \CC:\BE = \frac{1}{2} \CC: \left(\BF_{\rm e}\BF_{\rm e}^{\rm T} - \BI \right)
	\label{eq:S}
\end{equation}                                     
where $\CC$ is the anisotropic elastic stiffness tensor, $\BI$ is the second-order unit tensor, and $\BF_{\rm e}$  is an elastic deformation gradient which comprises elastic lattice stretch and rigid-body rotation. The deformation gradient $\BF$ derives from elastic and plastic deformations via
\begin{equation}
\BF= \BF_{\rm e} \BF_{\rm p}\quad,\quad           \BF_{\rm e} = \BF\BF_{\rm p}^{-1}
\label{eq:Fp}
\end{equation}
where the plastic deformation gradient $\BF_{\rm p}$ describes deformation due to  dislocation slip, mapping a vector in the reference configuration to the intermediate configuration. For a given deformation gradient, the partitioning between elastic and plastic deformation gradients determines the stress response, and the proportion of plastic deformation will increase once the material yields. The evolution of the plastic deformation gradient is determined by the plastic velocity gradient $\BL_{\rm p}$,
\begin{equation}
	\dot{\BF}_{\rm p} = \BL_{\rm p}\BF_{\rm p}\quad,\quad \BL_{\rm p} = \dot{\BF}_{\rm p}\BF_{\rm p}^{-1}
\end{equation}
Following the approach of Kalidindi \cite{kalidindi1998incorporation}, $\BL_{\rm p}$ in the intermediate configuration is determined by summing the plastic contributions from dislocation slip in each slip system,
\begin{equation}
	\dot{\BL}_{\rm p} = \sum_{\varsigma} \dot{\gamma}^{\varsigma} \Bn^{\varsigma}\otimes \Bs^{\varsigma} \quad,\quad
	\dot{\gamma}^{\varsigma} = \rho^{\varsigma} b^{\varsigma} v^{\varsigma}.
\end{equation}
As dislocation motion is controlled by an effective shear stress, the plastic velocity gradient depends on the stress state as well as the material state. Additionally, the evolution rate of the material state variables, such as dislocation density, is also related to the stress state and material state. Therefore, the update procedures for stress and material state have to be considered simultaneously. 

\subsection{Time integration}
In order to find a solution with consistent stress and material state, an implicit Euler scheme, i.e., fixed-point iteration, is adopted. Specifically, the integration of stress is performed at constant material state, and the integration of material state is then performed for a given stress until the residuals satisfy a convergence criterion.

\subsubsection{Stress integration}
Eq. (\ref{eq:Fp}) can be expressed in an implicit manner at fixed material state:
\begin{equation}
\frac{\BF_{\rm p}(t_i)-\BF_{\rm p}(t_{i+1})}{\Delta t}=\BL_{\rm p}(t_i+1)\BF_{\rm p} (t_i)   
\end{equation}                                         
where $t_i$ is the time at the beginning of the increment and $t_{i+1}$ is the time at the end of the increment. When a small time increment is used, the plastic and elastic deformation gradient are approximately given by
\begin{equation}
	\BF_{\rm p}(t_{i+1}) = (\BI - \Delta t \BL_{\rm p}(t_{i+1}))^{-1}\BF_{\rm p}(t_i)\quad,\quad \BF_{\rm e}(t_{i+1}) =\BF(t_{i+1}) \BF_{\rm p}(t_{i+1})^{-1}.
\end{equation}                                          
The second Piola–Kirchhoff stress tensor at the time $t_{i+1}$ is then computed from Eq. (\ref{eq:S}).
                                 
\subsection{Material state integration}
The material state variables in our model include the total dislocation densities $\rho^{\varsigma}$ on each slip system, as well as the dislocation density vectors $\Brho^{\varsigma}$ which contain information about edge and screw components of the geometrically necessary dislocation densities, dislocation velocities $v^{\varsigma}$, back stresses $\tau_{\rm b}^{\varsigma}$, and friction-like stress $\tau_{\rm b}^{\varsigma}$. According to Eqs. (\ref{eq:Brho}) and (\ref{eq:rho}), the evolution of the dislocation densities for a given stress depends on the values of the material state variables and on their spatial derivatives. 

Since dislocation transmission at grain boundaries is not considered, transport occurs within the individual grain onlys. The size, shape, and position information of each grain are stored during initialization of the simulation. For the evaluation of transport terms and nonlocal stress contributions we use a semi-implicit scheme where these terms are evaluated based on the constitutive response at the previous converged time step, and the post-update nonlocal variables then affect the evolution of stress and state at current time step. Local terms in the evolution equations, on the other hand, are calcuated using an implicit update:
\begin{equation}
	\rho^{\varsigma}(t_{i+1})= (\rho^{\varsigma}(t_i) + \Delta t \dot{\Brho}^{\varsigma}_{\rm NL} (t_{i})) + \Delta t \dot{\rho}^{\varsigma}_{\rm L} (t_{i+1})                        
\end{equation}                                     
 Spatial derivatives of field variables are calculated in a computationally efficient manner by double Fourier transform. We denote the Fourier transform and its inverse as $\cal F$ and ${\cal F}^{-1}$, with the understanding that these operations are performed on a regular periodically continued grid. The gradient operation of a scalar field $\gamma(\Br)$ is expressed in Fourier space as
\begin{equation}
\gamma(\Bk) = {\cal F}[\gamma(\Br)]     \quad,\quad                                                 
\nabla \gamma(\Br) = {\cal F}^{-1}[i\Bk \gamma(\Bk)] 
\end{equation}
where $\Br$ denotes the position vector of integration points in 3D, the vector $\Bk$ the wave vector in Fourier space, and $i$ the imaginary unit. Since the shear strain and strain rate fields are defined within individual grains only, they exhibit of necessity discontinuities at grain boundaries. Therefore, some care must be taken to avoid spurious short-wavelength oscillations on the characteristic length scale of the grid spacing that may result from the discrete FFT. To this end, a Gaussian low-pass filter is used when evaluating the derivatives of the plastic strain rate:
\begin{equation}
	h(\Bk) = \exp\left[-\frac{\Bk^2}{k_0^2}\right]
\end{equation}                                                
The gradient of a field state is then evaluated as
\begin{equation}
\nabla \gamma(\Br) = {\cal F}^{-1}[i\Bk h(\Bk) \gamma(\Bk)]                                         
\end{equation}
The Gaussian low-pass filter is tantamount to convoluting the field with a Gaussian coarse graining function. We choose the standard deviation $\lambda$ of this function according to the pragmatic criteron that (a) it must be sufficiently large to suppress spurious oscillations arising from the double FFT on a discrete grid, and (b) the constitutive response of the material must not depend on the numerical parameter $\lambda$. The value used throughout the following simulations is $\lambda = 5.4 l$ where $l$ is the grid spacing. 

\section{Results for Magnesium bi- and polycrystals}

We apply our simulation framework to deformation of polycrystalline Mg. We first investigate for an idealized multilayer geometry (a periodically continued bicrystal) the distribution of geometrically necessary dislocations, the formation of dislocation pile ups at grain boundaries, and the concomitant development of back stresses. We then move to a random Mg polycrystal created via Voronoi tesselation and investigate how these features translate into a grain size dependence of the rate of dislocation accumulation, and into Hall-Petch like behavior of the flow stress. 

We focus on deformation of ultrafine-grained systems, hence, we do not consider twinning as a dominant deformation mechanism \cite{li2011effects}. Deformation occurs mostly by basal slip, with prismatic or pyramidal slip systems playing a secondary role. The hardening matrix is taken from the work of Bertin et.al. \cite{bertin2014strength} and given in appendix A (Table A1). In Appendix A, we also provide a list of all parameters of our CP model as used in the present computation. 

\subsection{Plasticity boundary layers in a Mg multilayer}

\begin{figure}[tb]
	\includegraphics[width = 0.8\textwidth]{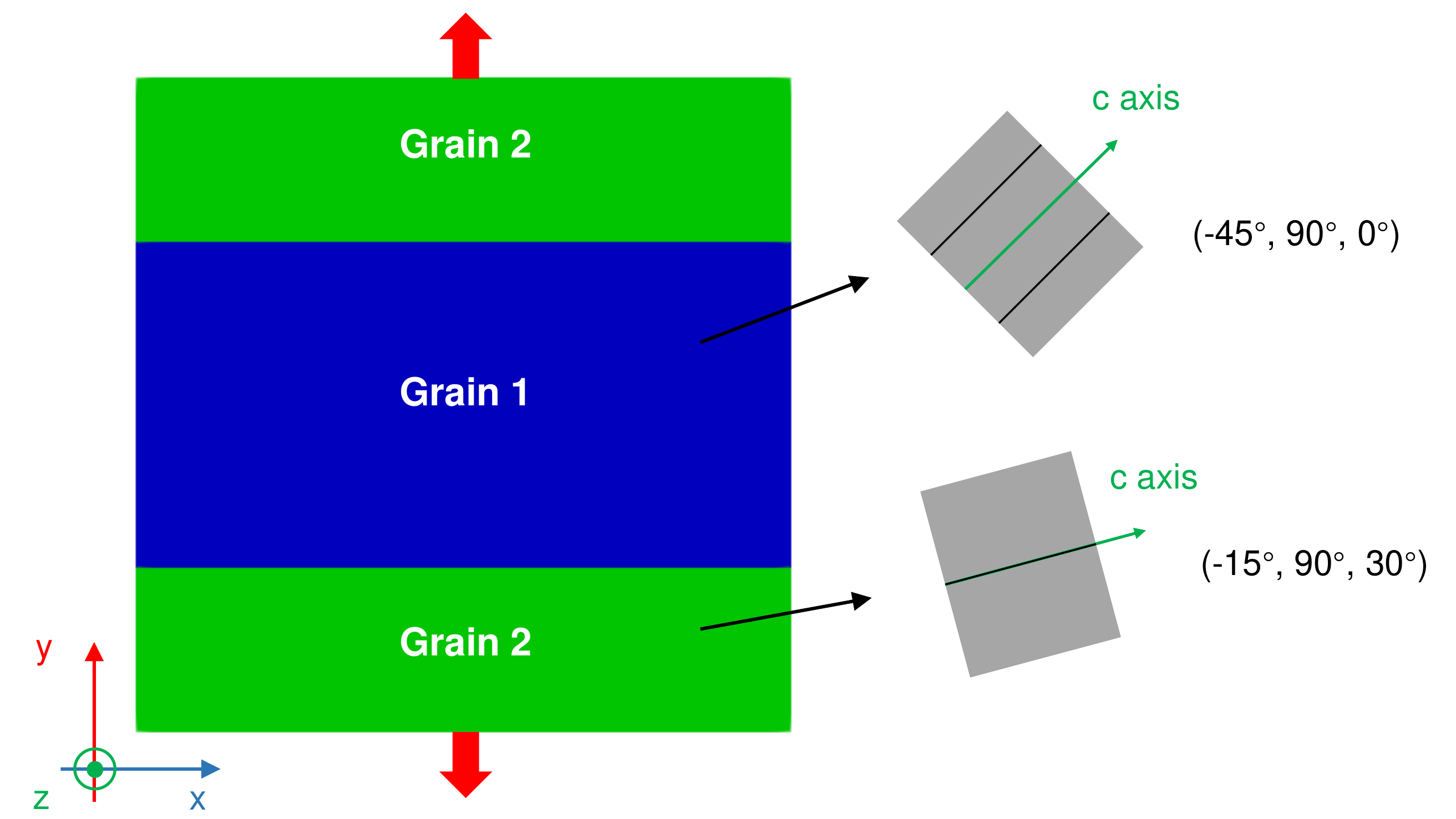}
	\caption{Schematic illustration of our bicrystal model, indicating grain orientations and showing the local
		coordinate systems in the crystallites. } 
	\label{fig:bicrystal}
\end{figure}

We consider a bicrystal consisting of two planar Mg grains with orientations shown in \figref{fig:bicrystal}. While one grain (here placed in the center) deforms almost exclusively by basal slip, the second (here:outer) grain is unfavorably oriented for basal dislocation motion. Since we impose periodical boundary conditions in all directions, the system can also be envisaged as a multilayer consisting of grains of alternating 'soft' and 'hard' orientation. The multilayer is loaded in tension in the direction perpendicular to the layers ($y$ direction of the global coordinate system). 
Because of the symmetry of the problem, deformation occurs in plane strain, and all components of strain in the $xy$ plane as well as all internal variables can be expressed as functions of the global $y$ coordinate only. For the same reason, it is straightforward to perform spatial averages, which can be written in terms of integrals over the global $y$ coordinate, and certain average quantities can be evaluated analytically. 

\begin{figure}[tb]
	\includegraphics[width = 0.9\textwidth]{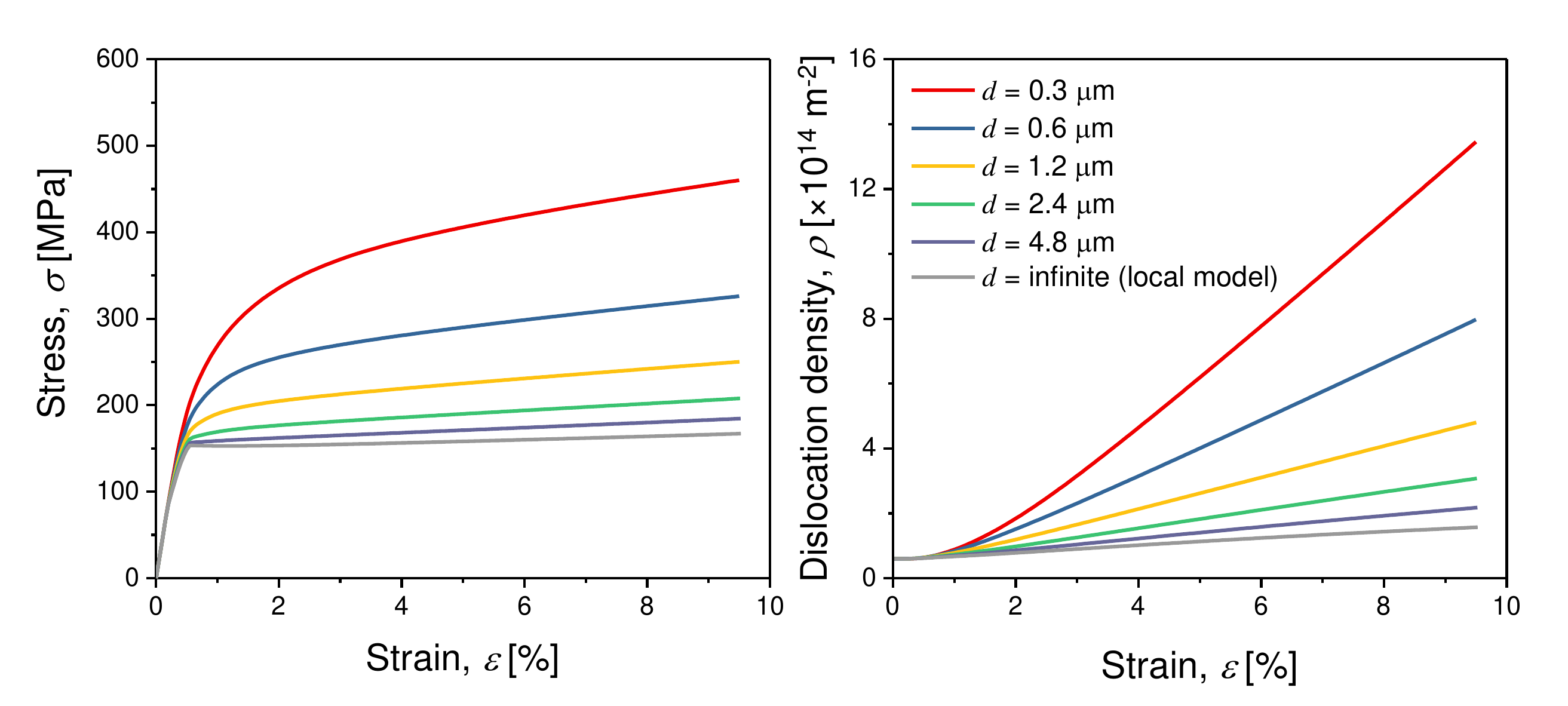}
	\caption{Left: Stress-strain curves of bicrystal multilayers for five different layer thicknesses; right: evolution of average dislocation density for the same samples. For reference, we also show results for a model without transport or back stress, corresponding to layer thicknesses that are so large that boundary effects near the layer boundary can be neglected. } 
	\label{fig:StressStrain} 
\end{figure}

\begin{figure}[tb]
	\includegraphics[width = 0.9\textwidth]{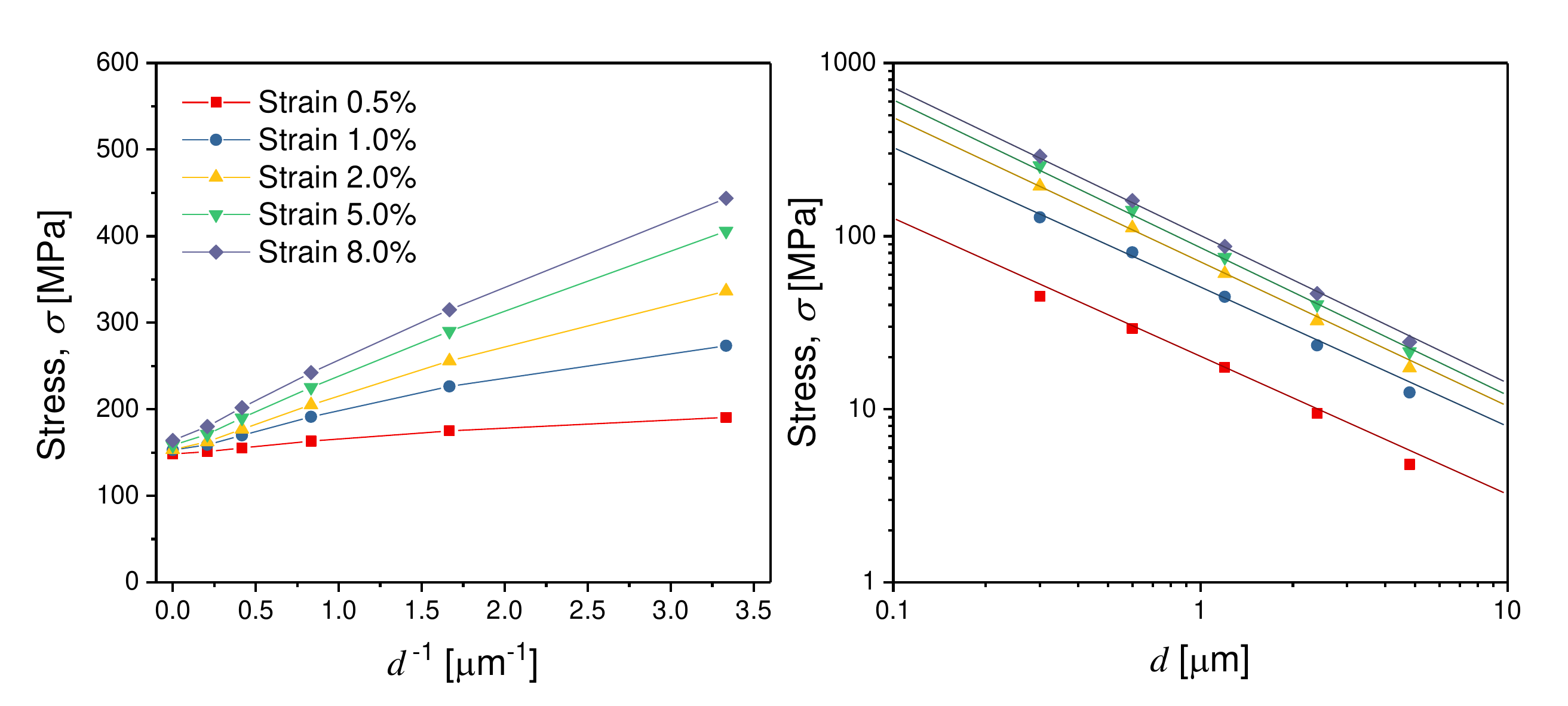}
	\caption{Dependency of flow stress on layer thickness for various strains. The slopes of the fits in the double-logarithmic plot on the right graph indicate a Hall-Petch exponent between $\delta = 0.8$ and $\delta = 0.85$. } 
	\label{fig:HallPetch}
\end{figure}

\figref{fig:StressStrain} shows simulated stress-strain curves for different layer thicknesses. In all simulations, constant and spatially homogeneous initial dislocation densities of $5 \times 10^{12}$ m$^{-2}$ were assigned to all slip systems, hence the initial conditions for the internal variables were $\rho^{\varsigma}_0 = 5 \times 10^{12}$ m$^{-2}, \; \Brho^{\varsigma}_0 = 0 \; \forall \varsigma$. An overview of model parameters is found in Appendix A. The simulated stres-strain curves exhibit a pronounced size effect: flow stress and dislocation accumulation rate both increase with decreasing layer thickness, as see in \figref{fig:StressStrain}. The flow stress $\sigma = \sigma_{yy}$ shows Hall-Petch-like behavior, 
\begin{equation}
	\sigma = \sigma_0 + \frac{K}{d^{\delta}}
    \label{eq:HP}
\end{equation}
where $\sigma_0$ depends both on the critical resolved shear stresses and the initial dislocation densities $\rho^{\varsigma}_0$ on the different slip systems. The Hall-Petch exponent is in our simulations close to $\delta = 1$ (\figref{fig:HallPetch}). 


\begin{figure}[tb]
	\includegraphics[width = 0.98\textwidth]{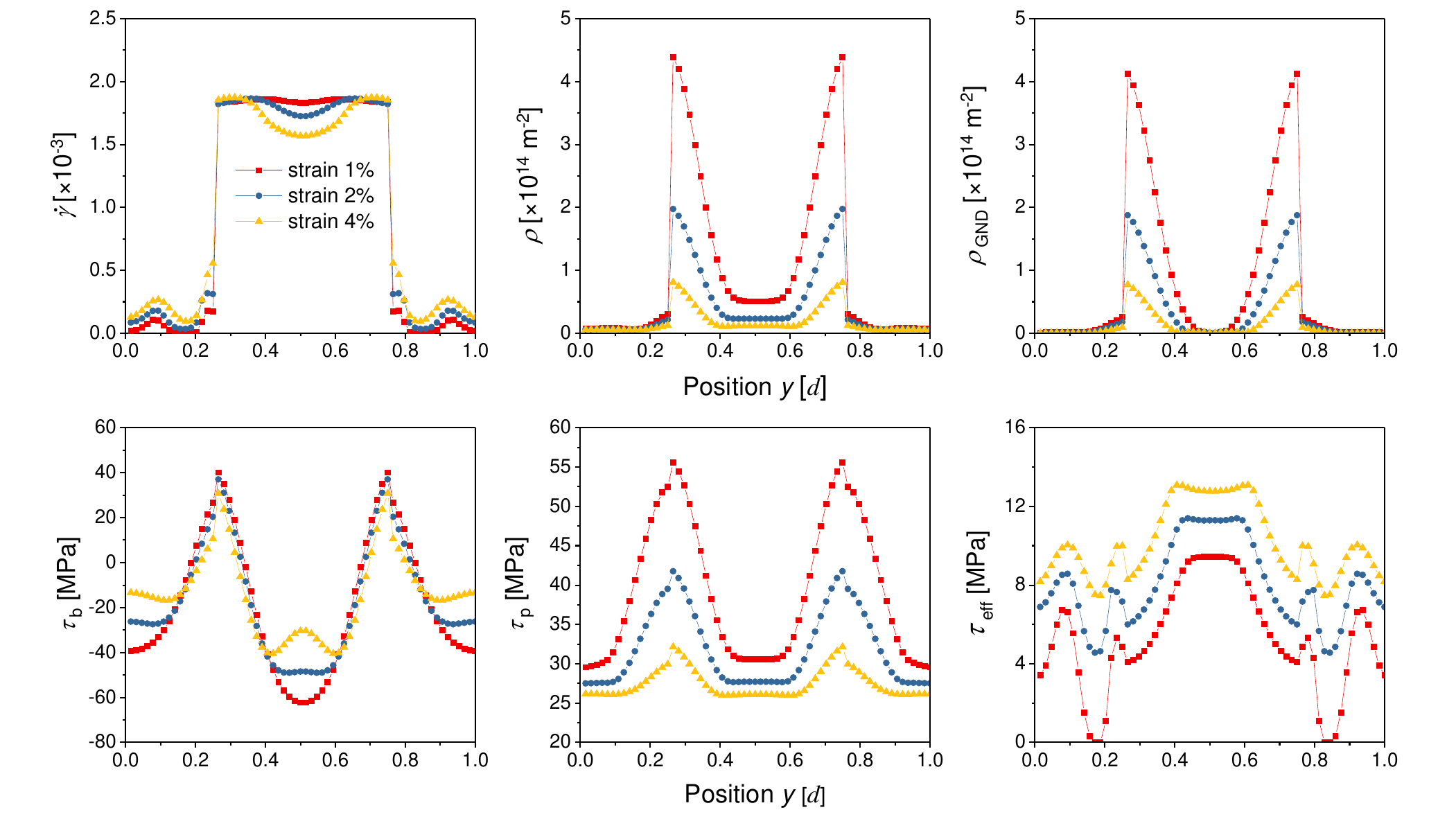}
	\caption{Profiles of internal variables at three different global strains, layer thickness $d = 1.2 \mu$m, showing the build-up of GNDs at the grain boundary and concomitant back stress $\tau_{\rm b}$ and enhanced friction stress $\tau_{\rm f}$; all variables refer to the basal slip system with the highest Schmid factor, from top left to bottom right: shear strain rate, total dislocation density, GND density, back stress, friction stress, effective resolved shear stress.}
	\label{fig:IntVarStrain}
\end{figure}

\begin{figure}[tb]
	\includegraphics[width = 0.98\textwidth]{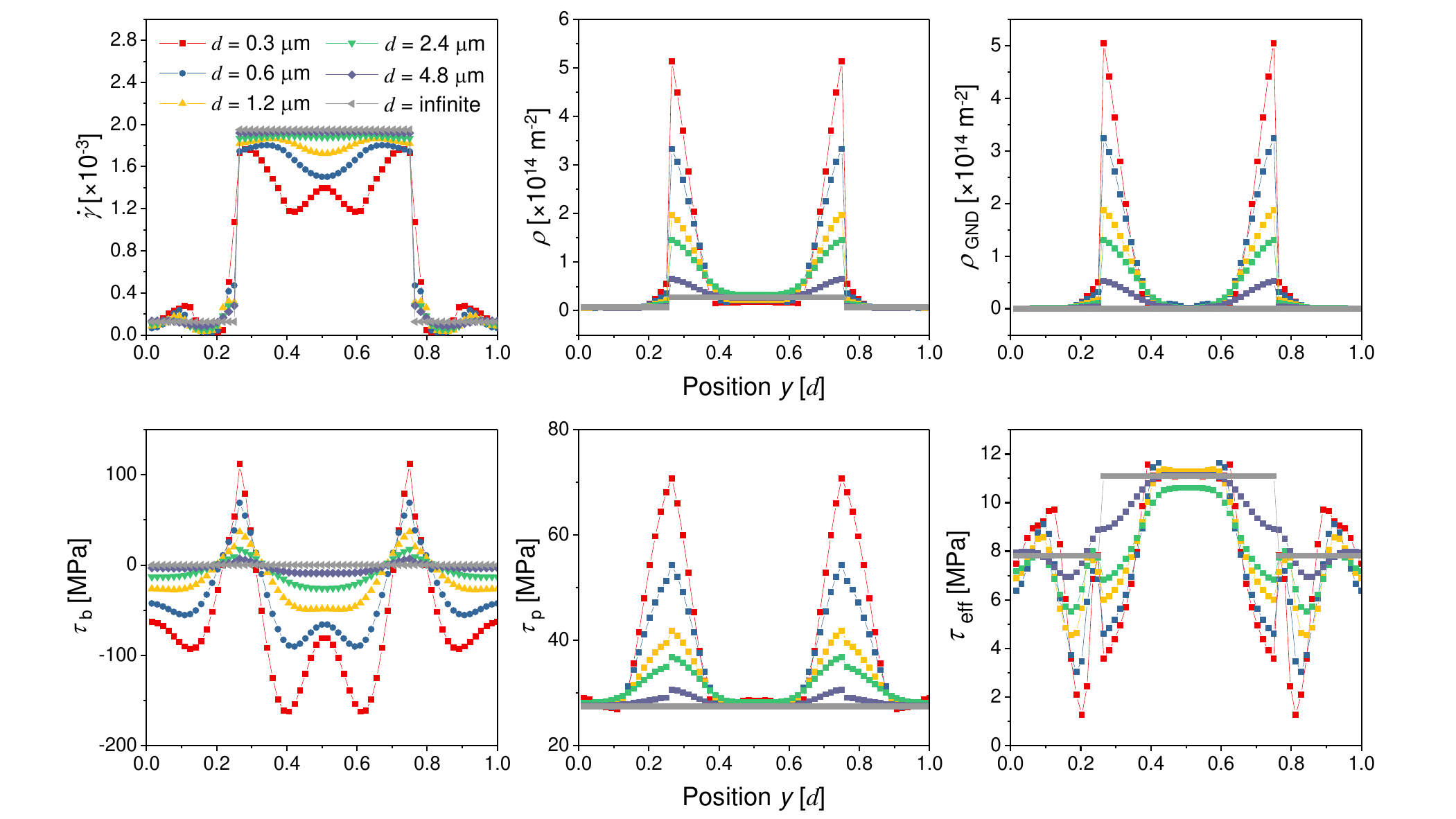}
	\caption{Profiles of internal variables for five different grain sizes, profiles in a simulation without transport and back stress are given for reference (grey curves), axial strain $\epsilon_{yy} = 2\%$; variables as in \figref{fig:IntVarStrain}.} 
	\label{fig:IntVarSize}
\end{figure}

To understand the reasons for this behavior we study the distribution of internal variables as shown in \figref{fig:IntVarStrain} and \figref{fig:IntVarSize}. During deformation, dislocations move from the grain interior to the grain boundaries where they pile up as geometrically necessary dislocations. After a short transient, practically all dislocations near the grain boundary are of GND type, i.e., $\rho^{\varsigma}_{\rm GND}/\rho^{\varsigma} \approx 1$. The opposite situation is found in the grain center where, for symmetry reasons, the strain gradient and thus the GND density must vanish. We are thus in a position to evaluate the average of the back stress (Eq. (\ref{eq:taub})), as follows: 
\begin{equation}
	\frac{2}{d} \int_0^{d/2}\tau_{\rm b}(y) \dy = \frac{2 \mu \eta_{\rm D} b^{\varsigma}}{d} \int_0^{d/2}\partial_y \left(\frac{\rho^{\varsigma}_{\rm GND}}{\rho^{\varsigma}} \right)\dy = 
	\frac{2 \mu \eta_{\rm D} b^{\varsigma}}{d}
\end{equation}
Thus, the average back stress is expected to obey Eq. (\ref{eq:HP}) with a Hall-Petch exponent $\delta = 1$. This stress is, after an initial transient required to create the dislocation pile-up at the grain boundaries, independent of strain, in agreement with the back stress profile evolution depicted in  \figref{fig:IntVarStrain}. We note that the finding that back stresses associated with dislocation pile ups leading to a Hall-Petch exponent close to 1 is consistent with the results of discrete dislocation dynamics simulations reported by Lu et. al. \cite{lu2022size}.

To understand the size dependent hardening, we start from Eq. (\ref{eq:qav}) and note that the  additional dislocation multiplication rate associated with the GND curvature can for the bicrystal geometry be expressed as
\begin{equation}
	\partial_t \rho^{\varsigma}_{\rm GND} = q^{\varsigma}_{\rm GND}v^{\varsigma} = 
	\frac{\dot{\gamma}^{\varsigma}}{b^{\varsigma}}\partial_y \left(\frac{\rho^{\varsigma}_{\rm GND}}{\rho^{\varsigma}}\right)
\end{equation}
We now average this expression over the interval between grain center and grain boundary and use the same argument as above for the back stress. This results in  
\begin{equation}
	\langle \partial_t \rho^{\varsigma}_{\rm GND}\rangle = 
	\frac{2\dot{\gamma}^{\varsigma}}{b^{\varsigma}d}
\end{equation}
Thus, the excess density is $\rho^{\varsigma}_{\rm GND} = 2\gamma^{\varsigma}/(b^{\varsigma}d)$. For the bilayer, an alternative consideration which does not rely on the mathematical framework of continuum dislocation dynamics is based on the expansion of loops of volume density $n^{\varsigma}$ which draw out geometrically necessary dislocations at the grain boundaries. The strain rate is in this case given by $\dot{\gamma}^{\varsigma} = n^{\varsigma} d b^{\varsigma} v^{\varsigma}$ and the rate of dislocation density increase is $\dot{\rho}^{\varsigma}_{\rm GND}= 2 n^{\varsigma} v^{\varsigma}$, which leads to the same result $\rho^{\varsigma}_{\rm GND} = 2\gamma^{\varsigma}/(b^{\varsigma}d)$. This consideration shows how the constraint imposed by the grain boundary leads to enhanced creation of dislocations, which is in the continuum dislocation dynamics formalism mediated by an enhanced curvature associated with the geometrically necessary dislocations. In conjunction with the Taylor relationship, the additional dislocations lead to an increased 'friction stress' which scales like $\tau_{\rm f} \propto \mu (\varepsilon^{\rm p} b/d)^{1/2}$. This line of reasoning indicates a parabolic shape of the stress-strain curves, consistent with \figref{fig:StressStrain}. Earlier qualitative arguments which follow the same line of reasoning can be traced back to Ashby \cite{ashby1970deformation}. For the Hall-Petch-exponent, the superposition of back stress hardening and enhanced friction stress leads to an effective Hall-Petch exponent slightly less than $\delta = 1$, as seen in \figref{fig:HallPetch}, right, where fits to the double logarithmic plots of flow stress vs. grain size produce values of $\delta \approx 0.8$.

\subsection{Uniaxial deformation of random Mg polycrystals}

\begin{figure}[tb]
	\includegraphics[width = 0.9\textwidth]{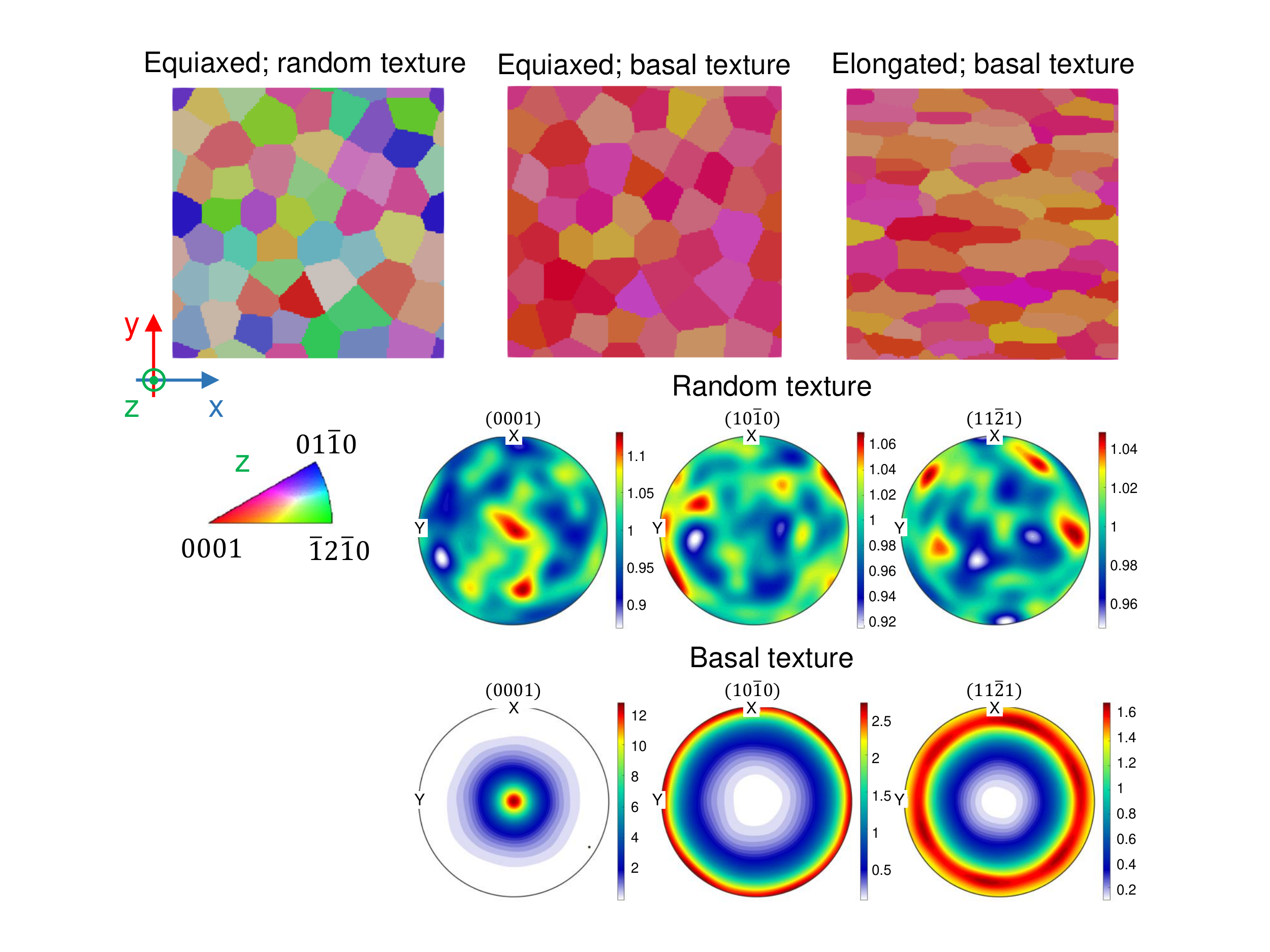}
	\caption{Polycrystal models considered in the simulations; all models use periodic boundary conditions with columnar grains in $z$ direction.} 
	\label{fig:poly}
\end{figure}

Moving to polycrystals, we consider three types of samples as shown in \figref{fig:poly}. All samples represent columnar quasi-two-dimensional grain structures (i.e., the grain morphology is simply continued in $z$ direction). Periodic boundary conditions are imposed in all three spatial dimensions.  Three different grain morphologies are considered: (i) equiaxed grains are defined using a Voronoi construction which is regularized by imposing a minimum seed distance of $0.7 d$ where $d$ is the mean grain size (see below), and grain orientations are assigned independently which are equidistributed over the unit sphere ('random texture'); (ii) a similar pattern of equiaxed grains is assigned grain orientations such that the $c$ axis directions are strongly clustered near the tensile axis ('basal texture'); (iii) the grain morphology is subjected to an affine stretch that induces an aspect ratio of 3:1 ('elongated grains'), also in this case a basal texture is used. The different models are illustrated in \figref{fig:poly}. or the elongated grain structure we consider two different loading directions, where the loading axis coincides either with the long or the short axis of the grains. 

The default model uses a grid of $128 \times 128$ lattice points accomodating $N = 50$ grains where $N$ is the number of seeds used in the Voronoi construction. The average grain size $d$ is then defined via $L^2 = N d^2$ where $L$ is the physical size of the model. The grain size is varied by scaling $L$, hence, the spatial resolution at which the grain microstructure is captured is always the same.

\subsubsection{Grain size dependence of deformation properties}

\begin{figure}[tb]
	\includegraphics[width = 0.95\textwidth]{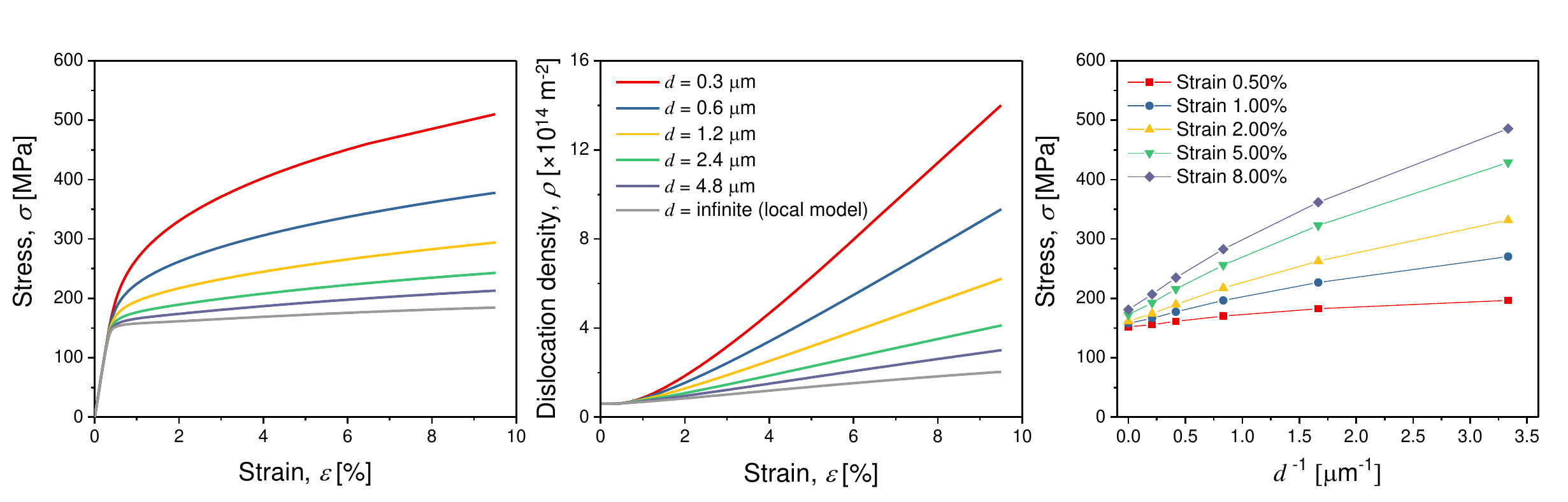}
	\caption{Left: Stress-strain curves of polycrystals of different grain size, all simulations consider equi-axed grains with basal texture as described in \figref{fig:poly}; center: dislocation density as function of strain; right: dependency of flow stress on grain size. For reference, we also show results for a model without dislocation transport or back stress. } 
	\label{fig:StressStrainPoly}
\end{figure}

We first investigate the grain size dependence of the deformation behavior, considering equi-axed grains with basal texture. The findings for polycrystals are very similar to those of the bicrystal multilayer system, as can be seen by comparing \figref{fig:StressStrainPoly} with \figref{fig:StressStrain}. In both cases, we observe Hall-Petch behavior with an exponent close to $\delta \approx 0.8$. In fact, the only quantitative difference to the multilayer system consists in a slightly higher dislocation density accumulation rate, and a correspondingly higher hardening rate, for the polycrystal. This leads to slightly higher values of the dislocation densities and flow stresses.

\subsubsection{Effects of grain morphology and texture}

\begin{figure}[tb]
	\includegraphics[width = 0.95\textwidth]{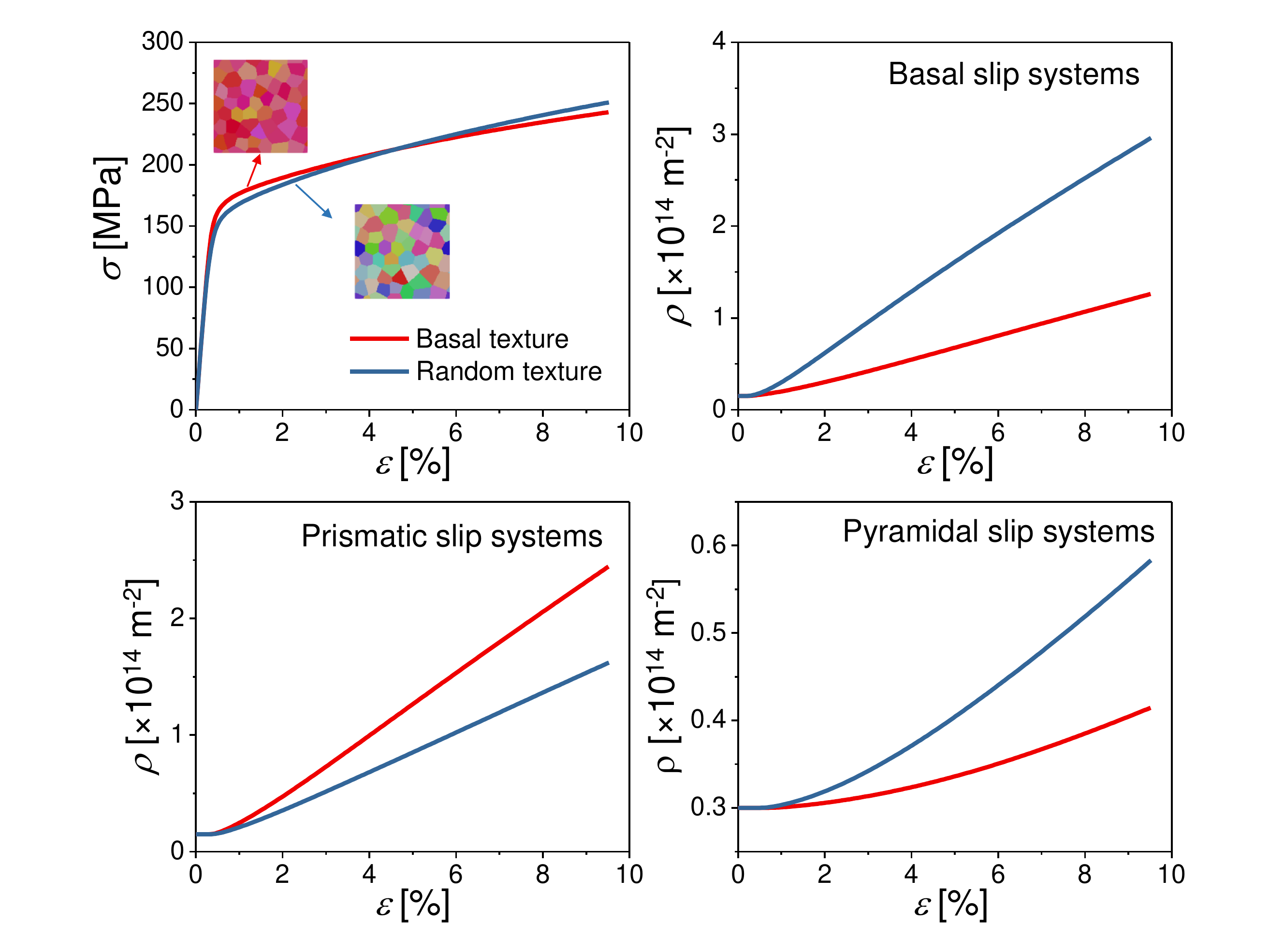}
	\caption{Stress-strain curves of polycrystals with equi-axed grains, with and without texture, and dislocation accumulation on the different types of slip systems; all samples with average grain size $d = 2.4 \mu$m. } 
	\label{fig:Texture}
\end{figure}

Next, we investigate the dependency of the macroscopic stress-strain curves on texture by comparing deformation of microstructures with equi-axed grains exhibiting either random or basal texture (\figref{fig:Texture}). The stress-strain curves shown in \figref{fig:Texture} are in agreement with the general idea that basal texture is detrimental to the overall deformation properties as it increases the yield stress and simultaneously reduces the hardening capability of the material. Basal texture also impacts the contribution of the different slip systems to the increase of dislocation density: Dislocation multiplication on the basal slip systems is suppressed, and this is compensated by enhanced dislocation multiplication on the prismatic systems. Deformation activity on pyramidal systems is low in both cases, though slighly higher in case of the randomly textured polycrystals. 

\begin{figure}[tb]
	\includegraphics[width = 0.95\textwidth]{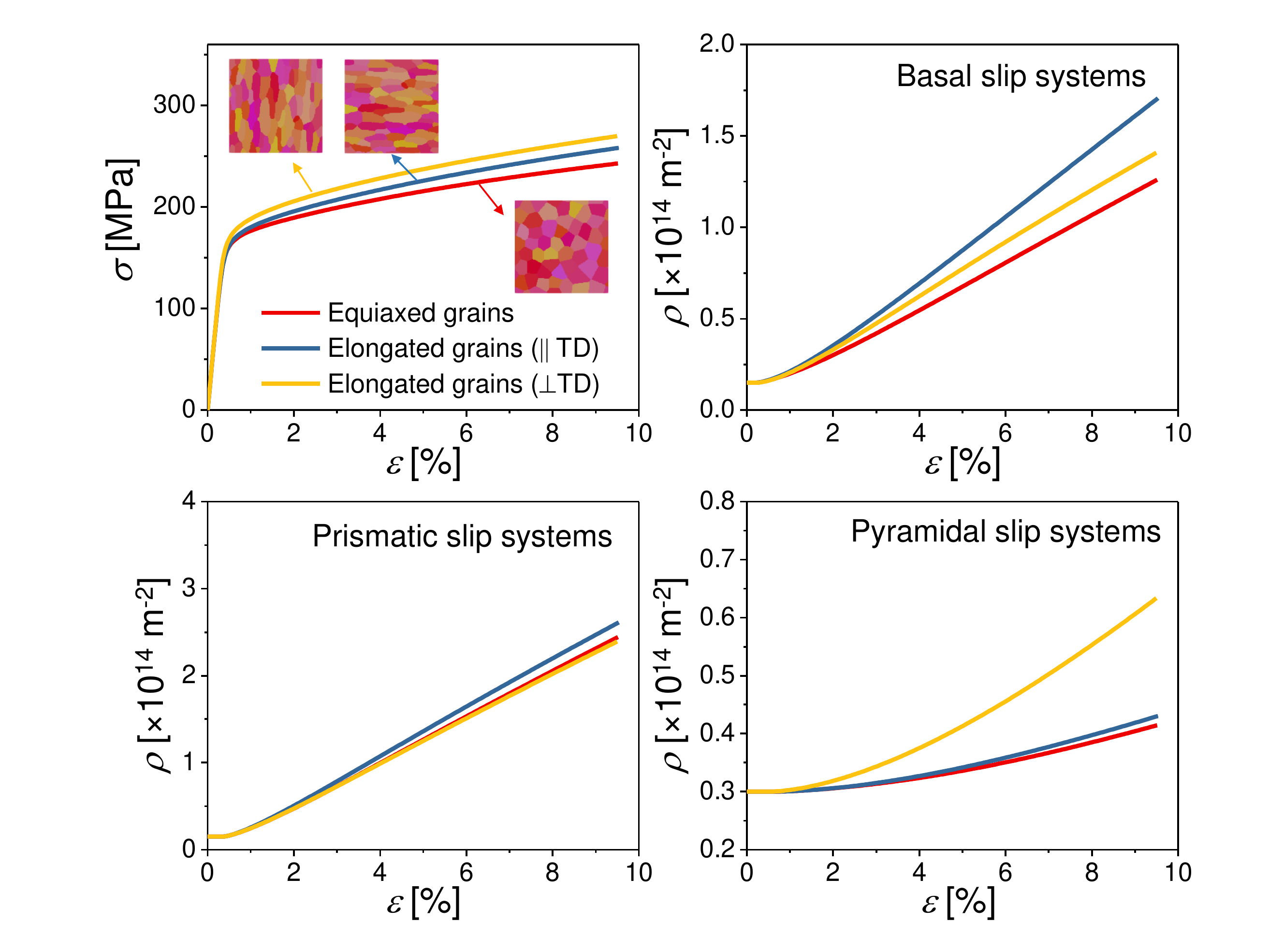}
	\caption{Stress-strain curves of polycrystals with basal texture, considering effects of grain morphology, and dislocation accumulation on the different types of slip systems; all samples with average grain size $d = 2.4 \mu$m. } 
	\label{fig:Morphology}
\end{figure}

Turning to effects of grain morphology, we observein \figref{fig:Morphology} that equi-axed grains have a lower flow stresses than elongated grains of the same $d$, i.e., the same average area per grain. Together with the observations on bicrystal multilayers, where the lateral extension of the grains is infinite, this observation indicates that the deformation properties are controlled by the length of the smaller half-axis of the ellipsoid that represents the average grain scale. In our simulation, aligning the tensile axis with the  long axis of the grains produced a slightly lower flow stress concomitant with enhanced activity on the basal slip systems; this observation is due to the specific orientation of the grains in that particular realization of the polycrystal model and does not represent a systematic trend. 

\begin{figure}[tb]
	\includegraphics[width = 0.95\textwidth]{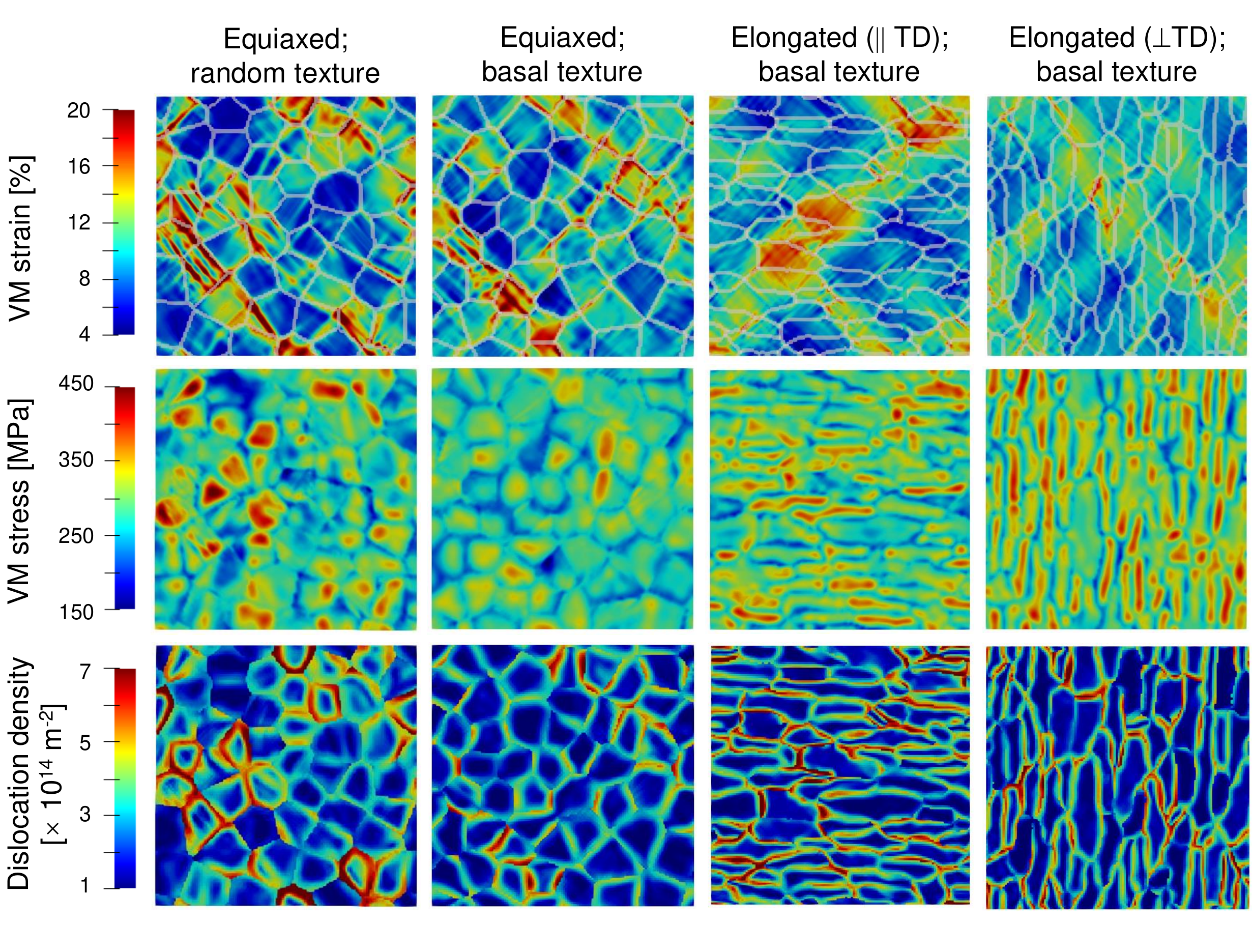}
	\caption{Spatial distribution of local equivalent stress, local equivalent strain, and dislocation density for different polycrystal models; all samples with average grain size $d = 2.4 \mu$m, total strain $\epsilon_{xx} = 10$\%. } 
	\label{fig:IntVarPoly}
\end{figure}

The spatial distribution of stress, plastic strain, and dislocation density for the different polycrystal models is illustrated in \figref{fig:IntVarPoly}, representing samples with average grain size $d = 2.4 \mu$m deformed to an axial strain of 10\%. In line with the observations on the bicrystal model, dislocations accumulate as GNDs at the grain boundaries. Dislocation pile-up is  strongest in grains where plastic activity on the basal plane is highest, it is therefore more pronounced in structures with random than in structuress with basal texture. We also note that dislocation pile up is stronger in structures with elongated grains, in line with the idea that this process is controlled by the shorter half-axis of the ellipsoid characterizing the average grain shape. 

Where grains of 'soft' orientation are aligned near $45^{\circ}$ to the tensile axis, the plastic strain distribution shows the formation of diffuse shear bands. In places, slip lines develop inside grains depending on local internal stresses, and merge across several grains. These localization features are more pronounced in equi-axed than in elongated grain structures, which exhibit a smaller effective grain size. 

Looking at the internal stress pattern, one observes that stresses tend to be higher in the grain interiors. This looks at first glance counter-intuitive but is readily understood from the fact that an enhanced stress level is required to maintain dislocation multiplication and plastic activity against the back stress created by the piled-up dislocations. 

\section{Summary and Conclusions}

We have developed a simplified continuum dislocation dynamics model which describes the generation, annihilation and transport of dislocations in a crystal plasticity framework. Compared to previous models, the present formulation avoids tracing explicitly the evolution of dislocation curvature. Instead of using a curvature evolution equation, which would necessitate handling higher-order spatial derivatives, curvature is here calculated from the basic density variables characterizing the dislocation system, namely the total dislocation densities on the different slip systems as well as the edge and screw contributions to the geometrically necessary dislocation density, which is represented as a dislocation density vector. Dislocation multiplication is described in terms of dislocation curvature, where on the one hand, the curvature of geometrically necessary dislocations is considered. This expression, which is exact in the limit where only GNDs are present, is complemented by a phenomenological assumption regarding the residual curvature which, in the limit where strain gradients are absent, reproduces well established phenomenological expressions for the dislocation multiplication rate. 

Thermodynamic consistency requires that dislocation multiplication must occur under a stress that is sufficiently high to ensure that the expended work matches the created defect energy, hence, the GND curvature 'attracts' a back stress term which depends on spatial derivatives of dislocation densities. 

We applied the framework to Mg multilayers consisting of alternating layers of 'hard' and 'soft' lattice orientation, as well as to Mg polycrystals. We showed that GND curvature influences the deformation behavior in two ways, first by increasing the net rate of dislocation accumulation and second by the back stress term. Both effects together give rise to a Hall-Petch like behavior with a Hall-Petch exponent $\delta \approx 0.8$ as often observed both in discrete dislocation simulations, and in experiments in the UFG regime. 

The present formulation is numerically efficient and allows to conduct crystal plasticity simulations with transport with very moderate computational cost. This makes it, after appropriate calibration of model parameters using experimental data, a promising tool for high-throughput calculations to optimize grain microstructures in view of grain size, grain morphology and texture such as to optimize materials properties \cite{bonfanti2021digital}.

\begin{backmatter}

\section*{Competing interests}
  The authors declare that they have no competing interests.

\section*{Author's contributions}
M.Z. formulated the model and prepared the manuscript, X.L. implemented the model and performed the simulations. Both authors edited the manuscript. 

\section*{Acknowledgements}
X.L. acknowledges financial support by CSC. 
. 


\bibliographystyle{bmc-mathphys} 
\bibliography{transport}      
\newpage\appendix
\section{Parameters of the crystal plasticity model for Mg}\label{app:param}

In our simulations of the crystal plasticity model, the following parameter values are used:\\[12pt]

{\em Material parameters}\\[6pt]
\begin{tabular}{lll}
	\hline
	Symbol & Meaning & Value and unit\\
	\hline
	$\mu$ & Shear modulus & 16.9 GPa \\
	$b_{\rm a}$ & Burgers vector length, basal and prismatic & 3.21 \AA\\
	$b_{\rm c+a}$ & Burgers vector length, pyramidal & 6.12 \AA\\
	\hline
\end{tabular}
\\[18pt]
{\em Deformation process parameters}\\[6pt]
\begin{tabular}{lll}
	\hline
	Symbol & Meaning & Value and unit\\
	\hline
	$\dot{\epsilon}_{yy, \rm ext}$ & Externally imposed axial strain rate & $0.001$s$^{-1}$ \\
	$T$ & Temperature in Kelvin & 300 K \\
	\hline
\end{tabular}
\\[18pt]
{\em Model parameters}\\[6pt]
\begin{tabular}{lll}
	\hline
	Symbol & Meaning & Value and unit\\
	\hline
	$\rho_0$ & Initial dislocation density, active slip systems & $5.0 \times 10^{12}$m$^{-2}$ \\
    $v_0$ & Reference velocity & $1.0 \times 10^{6}$ m/s \\
	$Q$ & Activation energy for dislocation motion & 0.85 eV\\
	$V_{\rm a}$ & Activation volume & 50$b^{3}$\\
    $\eta_{\rm D}$ & Factor controlling back stress& 5 \\ 
    $\eta_{\rm L}$ & Factor controlling dislocation generation& 0.08 \\
    $y_{\rm A}$ & Annihilation distance& 10$b$ \\
	\hline
\end{tabular}
\\[18pt]
{\em Slip system parameters}\\[6pt]
\begin{tabular}{lllll}
	\hline
	Number & Type & CRSS & Slip plane normal & Slip vector\\
	\hline
	B1 & & & (0 0 0 1) & [2 -1 -1 0]\\
	B2 & basal $\langle a \rangle$ & 1  MPa & (0 0 0 1) & [-1 2 -1 0]\\
	B3 & & & (0 0 0 1) & [-1 -1 2 0]\\
	\hline
	P1 & & & (0 -1 1 0) & [2 -1 -1 0]\\
	P2 & prismatic $\langle a \rangle$ & 40  MPa & (1 0 -1 0) & [-1 2 -1 0]\\
	P3 & & & (-1 1 0 0) & [-1 -1 2 0]\\
	\hline
    Pca1 & & & (-2 1 1 2) & [2 -1 -1 3]\\
    Pca2 & & & (1 -2 1 2) & [-1 2 -1 3]\\
    Pca3 & pyramidal $\langle a+c \rangle$ & 80 MPa & (1 1 -2 2) & [-1 -1 2 3]\\
    Pca4 & & & (2 -1 -1 2) & [-2 1 1 3]\\
    Pca5 & & & (-1 2 -1 2) & [1 -2 1 3]\\
    Pca6 & & & (-1 -1 2 2) & [1 1 -2 3]\\
    \hline
\end{tabular}
\\[18pt]
Interaction coefficients between the different types of slip systems are taken from the work of Bertin et. al. \cite{bertin2014strength}. We give them here for completeness together with the corresponding hardening matrix:\\[12pt]

{\em Interaction coefficients} $h_{\varsigma\varsigma'}$\\[6pt]
\begin{tabular}{lll}
	\hline
	Interaction & Designation & Value\\
	\hline
	S1 & Basal self-interaction & 0.150\\
	\hline
	S2 & Prismatic self-interaction & 0.150\\
	\hline
	S3 & Pyramidal self-interaction & 0.150\\
	\hline
	1 & Coplanar basal/basal & 0.150\\
	\hline
	2 & Prismatic/prismatic & 0.038\\
	\hline
	3 & Coplanar basal/prismatic & 0.707\\
	\hline
	4 & Non-collinear basal/prismatic & 0.054\\
	\hline
	5 & Collinear prismatic/basal & 0.535\\
	\hline
	6 & Non-collinear prismatic/basal & 0.060\\
	\hline
	7 & Semi-collinear basal/pyramidal & 0.367\\
	\hline
	8 & Non-collinear basal/pyramidal & 0.293\\
	\hline
	9 & Semi-collinear prismatic/pyramidal & 0.068\\
	\hline
	10 & Non-collinear prismatic/pyramidal & 0.088\\
	\hline
	11 & Semi-collinear pyramidal/basal & 0.017\\
	\hline
	12 & Non-collinear pyramidal/basal & 0.011\\
	\hline
	13 & Semi-collinear pyramidal/prismatic & 0.025\\
	\hline
	14 & Non-collinear pyramidal/prismatic & 0.015\\
	\hline
	15 & Semi-collinear pyramidal/pyramidal & 0.018\\
	\hline
	16 & Non-collinear pyramidal/pyramidal & 0.0042\\
	\hline
\end{tabular}
\\[18pt]

{\em Latent hardening matrix}\\[6pt]
\begin{tabular}{lllllllllllll}
	\hline
	Number & B1 & B2 & B3 & P1 & P2 & P3 & Pca1 & Pcs2 & Pca3 & Pca4 & Pcs5 & Pca6\\
	\hline
    B1 & S1 & 1 & 1 & 3 & 4 & 4 & 7 & 8 & 8 & 7 & 8 & 8\\
    \hline
    B2 & 1 & S1 & 1 & 4 & 3 & 4 & 8 & 7 & 8 & 8 & 7 & 8\\
    \hline
    B3 & 1 & 1 & S1 & 4 & 4 & 3 & 8 & 8 & 7 & 8 & 8 & 7\\
    \hline
    P1 & 5 & 6 & 6 & S2 & 2 & 2 & 9 & 10 & 10 & 9 & 10 & 10\\
    \hline
    P2 & 6 & 5 & 6 & 2 & S2 & 2 & 10 & 9 & 10 & 10 & 9 & 10\\
    \hline
    P3 & 6 & 6 & 5 & 2 & 2 & S2 & 10 & 10 & 9 & 10 & 10 & 9\\
    \hline
    Pca1 & 11 & 12 & 12 & 13 & 14 & 14 & S3 & 16 & 16 & 15 & 16 & 16\\
    \hline
    Pca2 & 12 & 11 & 12 & 14 & 13 & 14 & 16 & S3 & 16 & 16 & 15 & 16\\
    \hline
    Pca3 & 12 & 12 & 11 & 14 & 14 & 13 & 16 & 16 & S3 & 16 & 16 & 15\\
    \hline
    Pca4 & 11 & 12 & 12 & 13 & 14 & 14 & 15 & 16 & 16 & S3 & 16 & 16\\
    \hline
    Pca5 & 12 & 11 & 12 & 14 & 13 & 14 & 16 & 15 & 16 & 16 & S3 & 16\\
    \hline    
    Pca6 & 12 & 12 & 11 & 14 & 14 & 13 & 16 & 16 & 15 & 16 & 16 & S3\\
    \hline
\end{tabular}
\\[12pt]

\end{backmatter}
\end{document}